\documentclass[10pt]{article}

\usepackage{amsmath, amssymb, graphicx, parskip, hyperref}
\usepackage[a4paper]{geometry}

\title{Pattern Formation on Networks:\\
from Localised Activity to Turing Patterns}

\author{N.~J.~McCullen$^1$ and T.~Wagenknecht$^2$ (N.D.)}

\renewcommand{\epsilon}{\varepsilon}

\begin{document} 

\footnotetext[1]{n.mccullen@physics.org. Department of Architecture and Civil Engineering, University of Bath, UK.}
\footnotetext[2]{Formerly: Department of Applied Mathematics, University of Leeds, UK; now deceased.}

\maketitle

\begin{abstract}

Systems of dynamical interactions between competing species can be used to model many complex systems, and can be mathematically described by {\em random} networks.
Understanding how patterns of activity arise in such systems is important for understanding many natural phenomena. 
The emergence of patterns of activity on complex networks with reaction-diffusion dynamics on the nodes is studied here. 
The connection between solutions with a single activated node, which can bifurcate from an undifferentiated state, and the fully developed system-scale patterns are investigated computationally.
The different coexisting patterns of activity the network can exhibit are shown to be connected via a snaking type bifurcation structure, similar to those responsible for organising localised pattern formation in regular lattices.
These results reveal the origin of the multistable patterns found in systems organised on complex networks.
A key role is found to be played by nodes with so called {\em optimal degree}, on which the interaction between the reaction kinetics and the network structure organise the behaviour of the system.
A statistical representation of the density of solutions over the parameter space is used as a means to answer important questions about the number of accessible states that can be exhibited in systems with such a high degree of complexity.

\end{abstract}

\section{Introduction}

\subsection{Patterns on Networks}

Patterns arising through the interactions of individual system components are found throughout nature and society,  and much of science is dedicated to identifying and understanding the origin of such patterns.
Reaction-diffusion systems of partial differential equations (PDEs) were used by Alan Turing to explain chemical  pattern formation on a spatial domain \cite{turing1990chemical} and there has been recent interest in understanding the interactions in social systems from a physics perspective \cite{castellano2009statistical}.

Interactions between individual system components can be represented as networks \cite{strogatz2001exploring}, with network structure revealing much about the underlying systems and their dynamics \cite{palla2005uncovering}.
There is now growing interest in reaction-diffusion systems organised on complex network topologies, where the interactions are transmitted between nodes via the network links (or {\em edges}), particularly in systems with activator-inhibitor dynamics on the nodes.
Potential applications include pattern development from networks of activating and suppressing genes involved in embryonic development \cite{nakao2010turing}, as well as the evolution of complex structures ({\em autocatalytic sets}) in systems of competing proteins, such as could lead to the origin of life from a random starting condition \cite{jain1998autocatalytic}.
The networks connecting individuals in a social system also often have non-local connections \cite{watts1998collective} and non-trivial geometries \cite{erdos1960evolution, barabasi1999emergence}, making the concept of a {\em pattern} less obvious in such non-spatial domains.

Recent numerical results have revealed a multiplicity of bulk activation states ({\em Turing modes}) in a predator-prey type reaction-diffusion model on a scale-free ({\em Barab\'asi--Albert}) type network \cite{nakao2010turing}.
In addition, regular networks (lattices) -- as used to numerically investigate spatially extended systems of PDEs -- can exhibit localised patterns, which are connected via ``snaking bifurcations'' in the parameter space of the system.

This work demonstrates how solutions with a single activated node are connected to increasingly larger patterns of activity and bulk-modes via a similar  {\em snaking bifurcations} structure.
The importance of network structure (i.e.~the {\em node degree}) in the dynamics is also demonstrated.
These bifurcations lead to a highly complex {\em ``turmoil''} of coexisting states, necessitating a statistical approach to understanding the behaviour of the system.

\subsection{Pattern Formation and Reaction-Diffusion Systems}

Alan Turing laid down the basis for pattern formation  on a spatial domain, based the loss of stability of an unpatterned equilibrium to another non-trivial (patterned) state \cite{turing1990chemical}, in an attempt to explain morphogenesis in embryonic development.
Such situations can be set up as a system of competing chemical species in a reaction-diffusion system:
\begin{align} 
\begin{split}
\dot{u} & =  f(u,v) + \epsilon \nabla^2 u \\
\dot{v} & =  g(u,v) + \sigma \epsilon \nabla^2 v,
\end{split}
\end{align}
where $u$ and $v$ are activator and inhibitor chemical species, $f$ and $g$ are functions for the internal {\em reaction} component at any location and $\nabla^2$ is the diffusion operator.
Such formulations have been widely used to describe pattern formation in a wide variety of systems on a spatial domain.
In numerical investigations of such systems space is discretised, with the reactions taking place on nodes on a regular mesh and diffusion occurring to neighbouring nodes via local network edge connections on the lattice.
In these cases the diffusion operator is replaced with the Laplacian matrix $L$, which represents the difference terms in the system of equations:
\begin{equation} \label{e:rdco}
\begin{split} 
\dot{\bf u} & =  f({\bf u}, {\bf v}) + \epsilon L {\bf u},\\
\dot{\bf v} & =  g({\bf u}, {\bf v}) + \sigma \epsilon  L {\bf v}.
\end{split}
\end{equation}

\subsection{Diffusion on Complex Networks}

In contrast to the continuous case, diffusion in discrete models can also take place along the edges of a more irregular underlying network. 
If the network has $N$ nodes $n_1, n_2, \ldots n_N$, diffusion is mediated by the network Laplacian $L$, defined in the same way as the regular case above, such that $L=(L_{(i,j)})$, with $L_{(i,j)}=1$ if nodes $i$ and $j$ are connected, $L_{(i,j)}=0$ if they are not, and $L_{(i=j)}=-d_i$, where $d_i$ is the {\em degree} of node $i$, such that each row sums to zero. 
The matrix $L$ therefore describes diffusion in a system such that the flux to node $i$ is given by the term:
$\sum_{j=1}^N L_{(i,j)} u_j$.

Nakao and Mikhailov  studied the Turing instability in large scale-free ({\em Barab{\'a}si--Albert} \cite{barabasi1999emergence}) networks \cite{nakao2010turing}.
Their numerical investigations revealed the coexistence and multi-stability of a huge variety of patterns, as can be seen in figure 3 of their paper.
They also found that stable patterns can exist before the homogeneous equilibrium becomes unstable, in a {\em subcritical} bifurcation.

\subsection{The Role of Network Structure and the Optimal Degree}

Wolfrum analysed the above system by considering the stability of an individual node of degree $d_k$ using:
\begin{equation} \label{e:wolfrum}
\begin{split} 
\dot{ u_k} & =  f({ u_k}, { v_k}) + d_k\epsilon (\bar{u} - u_k),\\
\dot{ v_k} & =  g({ u_k}, { v_k}) + \sigma d_k\epsilon (\bar{ v}-v_k),
\end{split}
\end{equation}
where the other nodes are considered fixed at the equilibrium $(\bar u, \bar v)$ \cite{wolfrum2012turing}.
This analysis, similar to the mean-field approach used elsewhere \cite{pastor2001epidemic}, was used in order to explain the bifurcations of the most basic single differentiated node (SDN) states from the undifferentiated equilibrium-state.
A further consequence of this analysis is that the stability of \eqref{e:wolfrum} depends on the interaction between the reaction terms (contained in the functions $f$ and $g$) and the node degree.
This leads to the prediction that a certain {\em preferential degree} can exist, for which activated solutions appear at some minimum value of the control parameter $\sigma$.
This result reveals an important connection between network structure and the dynamics of the system.

\subsection{Localised Patterns and Snaking Bifurcations}

As well as Turing patterns covering the whole spatial domain, spatially localised patterns have also been observed in a variety of experimental and numerical systems.
In particular, localised buckling was observed in experiments on cylindrical shells by Hunt et al. \cite{hunt1999buckling}, and explained using numerical continuation using discretisation schemes. 
Numerical continuation (as well as certain well-controlled experiments) have shown that localised patterns of increasing spatial extent are connected by folds and cusps in the one-parameter bifurcation diagram, in what are called {\em ``snaking bifurcations''}.
Snaking bifurcations originate in a subcritical bifurcation from the equilibrium state and are so-called due to their structure then winding upwards back-and forth in the one-dimensional parameter space.
As the bifurcation curve snakes upwards in amplitude, the spatial pattern grows incrementally by one wavelength for each bifurcation curve, as summarised in \cite{hunt2006buckling}.
The origin and development of these localised solutions via snaking has now been extensively investigated in numerical and analytical studies of reaction--diffusion systems on regular lattices. 
The connection between localised patterns via snaking bifurcation has been shown to occur via a sequence of homoclinic folds in the codimension-2 bifurcation diagram \cite{champneys2012homoclinic}.

Other bifurcation structures can develop in these systems, due to symmetry-breaking perturbations to the system \cite{knobloch2011nonreversible, knobloch2012non}.
Examples of such structures include the splitting up of snakes into a rung-like structure (so called {\em ``snakes and ladders''}) \cite{knobloch2011isolas}, as well as self-connected solutions that are not linked to the ground-state ({\em``isolas''}) \cite{sandstede2012snakes}.
The mechanism for the transition between localised patterns via snaking bifurcations is therefore theoretically well established for such regular systems.

Aside from reaction-diffusion systems, similar localised patterns have also been observed in physical experiments and bifurcation studies on other systems. 
Examples include experimental observations in optical systems \cite{Barbay2008}, numerical results for plain Couette flow \cite{schneider2010snakes}, as well as experimentally and analytically for hexagonal localised patterns in magnetic fluids \cite{lloyd2015homoclinic}. 
Bifurcation analysis has shown that it is also a snaking bifurcation structure that connects the different localised patterns in these systems, indicating that this is a robust and universal mechanism.

 \section{Methods}

\subsection{The Mimura-Murray Model}

The current investigation focusses on the Mimura-Murray model of prey-predator populations \cite{mimura1978diffusive},  following on from the work of Nakao and Mikhailov \cite{nakao2010turing}.
The kinetics of the reaction-diffusion system (eqn.~\eqref{e:rdco}) are described by the following equations:
\begin{equation}
f(u,v) = \frac{au+bu^2-u^3}{c} - uv, \quad g(u,v) = uv-v -dv^2,\label{eqn:mimu}
\end{equation}
where $u$ denotes the activator (or prey) and $v$ the inhibitor (or predator).
In these investigations $\sigma$ is the bifurcation (control) parameter, related to the relative strengths of the diffusion terms.
The current study was carried out using the  values $a=35$, $b=16$, $c=9$, $d=2/5$, $\epsilon=0.12$, at which the system possess the equilibrium $(\bar{u}, \bar{v})=(5,10)$.
This {\em undifferentiated} state is stable for small values of $\sigma$ but loses stability in a {\em subcritical} bifurcation at $\sigma \approx 15.5$.
In continuous media this would result in the emergence of alternating activator-rich and activator-low domains (a periodic {\em Turing-type} pattern in the supercritical case), but organised on networks can also display small-scale (``localised'') patterns of node activity in these subcritical cases.
Using the current parameter values in the reaction functions \eqref{eqn:mimu}, linear stability analysis of equation \eqref{e:wolfrum} results in an optimal degree of $d_k^*=9$, which is shown in this work to play a key role in organising the behaviour of the system.

\subsection{Network Properties}

In order to investigate the development and growth of ``patterns'' of activity on non-regular network topologies, the system was set up with the reaction species (on the nodes) interacting via the edges of a  scale-free random ({\em Barab{\'a}si--Albert} \cite{barabasi1999emergence}) network, as used in previous investigations \cite{nakao2010turing}.

\begin{figure}[b]
\centering
(a)\includegraphics[width=0.45\textwidth]{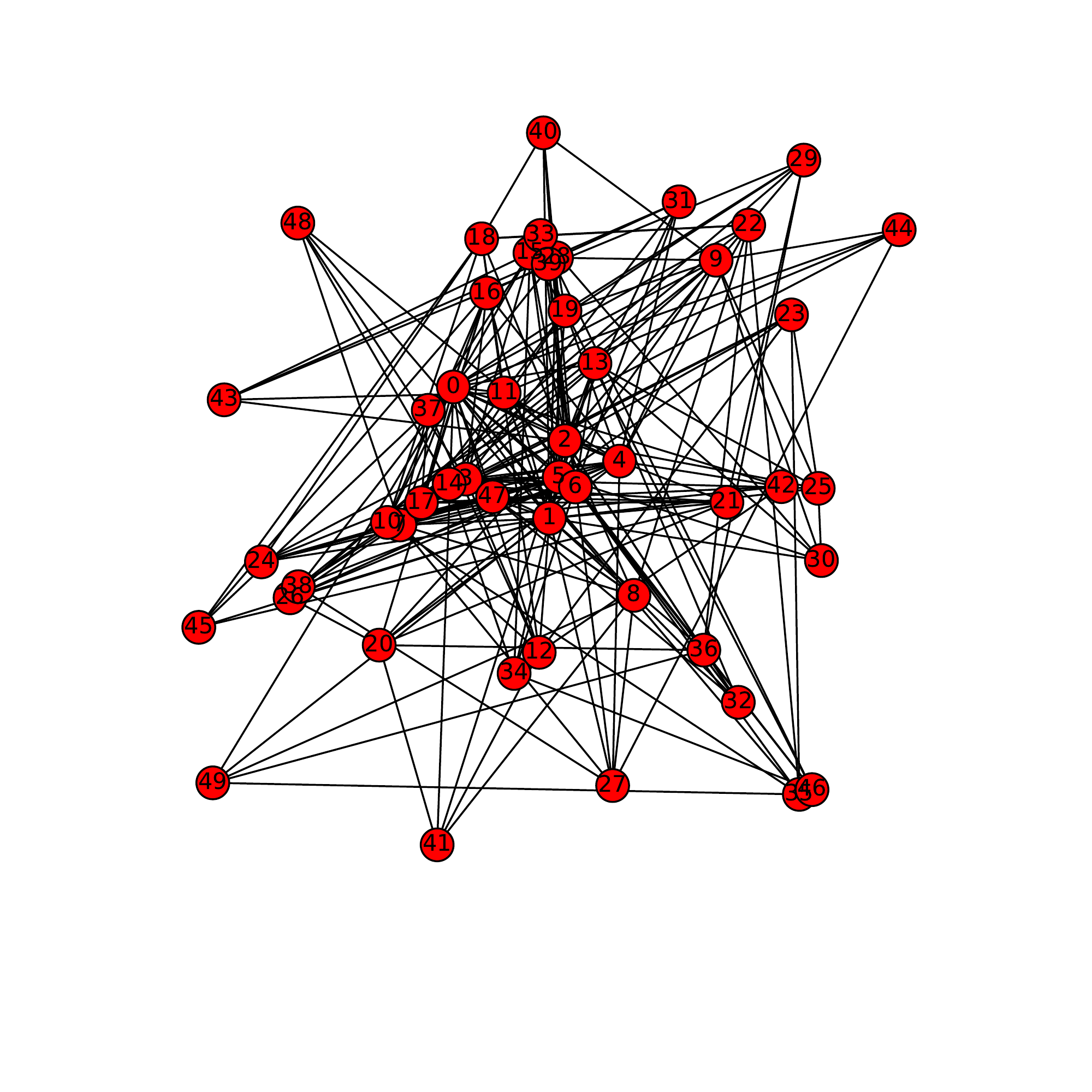}
(b)\includegraphics[width=0.45\textwidth]{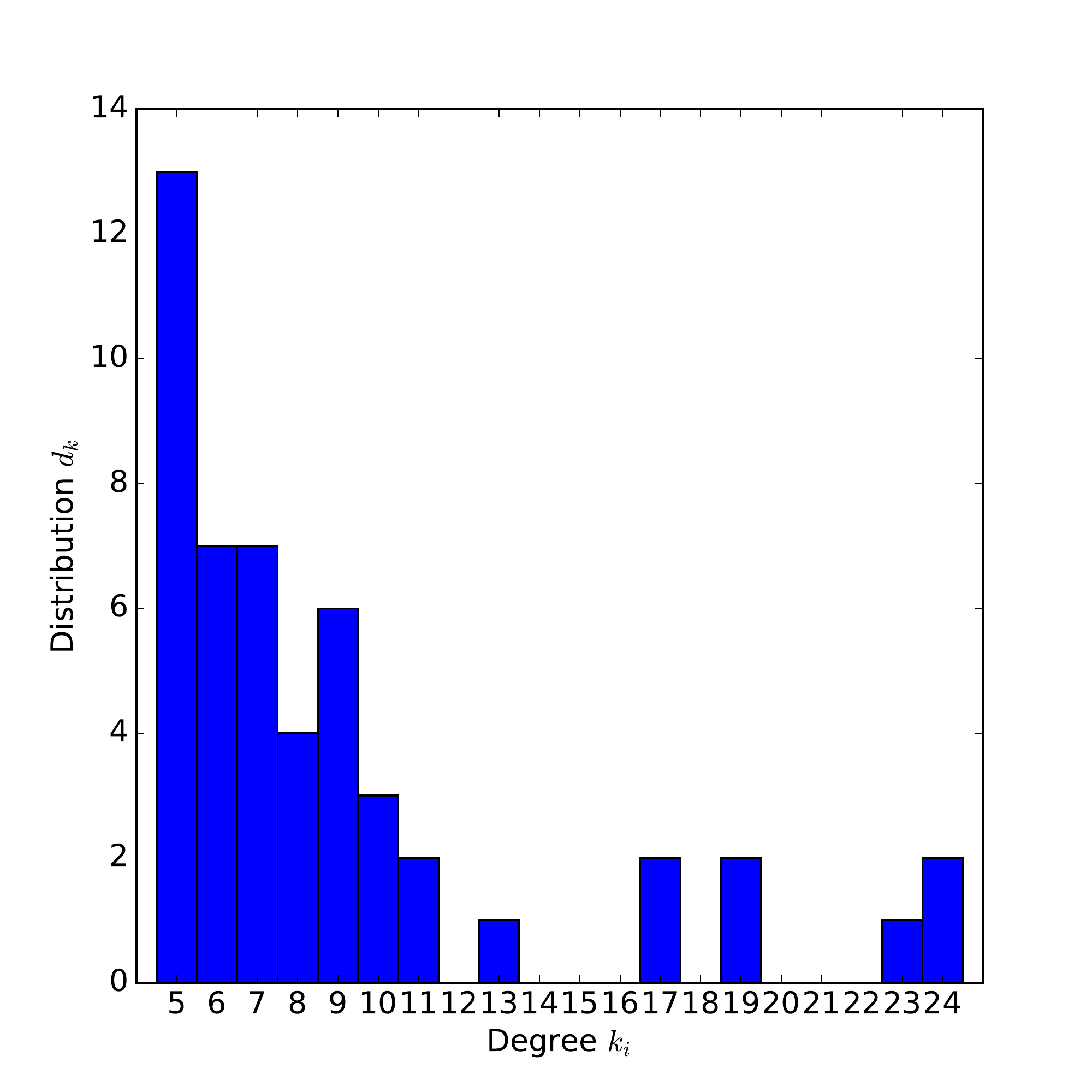} 
\caption{An example network used in this investigation (a), using the preferential attachment scheme of Barab{\'a}si and Albert.  The number of nodes is $N=50$ and degree of attachment for each newly added node is $M=5$. 
(b) This scheme gives a scale-free distribution in the limit of $N\rightarrow\infty$, but here, where $N$ is finite, the degree distribution only approximates scale-free.}\label{fig:network}
\end{figure}

A network with 50 nodes was used for these investigations, where new nodes added at each generation step have five edges assigned preferentially to higher degree existing nodes. 
The routines in the {\em NetworkX} module for Python were used to generate and visualise the networks.
Different topologies with the same characteristics were investigated to ensure consistent qualitative behaviour, but only one representative realisation is shown here.
For the majority of results in this work the {\em degree of attachment} $M=5$ was used in the network generation scheme (Fig.~\ref{fig:network}).
In these cases it can be seen that numerous nodes of the optimal $d_k^*=9$ lie towards the middle of the degree distribution.
For the later (comparison) cases $M=10$ was used to generate the network, in which the optimal degree nodes instead have amongst the lowest degree in the network (being close to the {\em periphery}), resulting in important differences in the system behaviour.

\subsection{Numerical Techniques}

Computations were started from the single-node (SDN) solutions, studied in previous work \cite{wolfrum2012turing}, which bifurcate from the undifferentiated state at the point $\sigma_T$. 
All possible singly differentiates node (SDN) solutions were found by numerically integrating the equations of the system from some random initial condition then refining using numerical root finding (using the widely available \texttt{fsolve} routine).
The solutions were then followed back and forth in their meander through the parameter space of the system using numerical {\em continuation} techniques provided by the AUTO bifurcation software \cite{doedel1981auto}.

\section{Results}

\subsection{From Localised Patterns to Large-Scale Activation}

Starting from single node (SDN) solutions and following the solutions in the bifurcation space  the transition between the SDN solutions and multiply differentiated node (MDN) states was investigated.
The connection between different states in the system was uncovered, as well as their coexistence with the fully developed (so called {\em ``Turing''}) patterns reported in \cite{nakao2010turing}.

\subsubsection{The growth of activation patterns.}

\begin{figure}
\centering
\includegraphics[width=.47\textwidth]{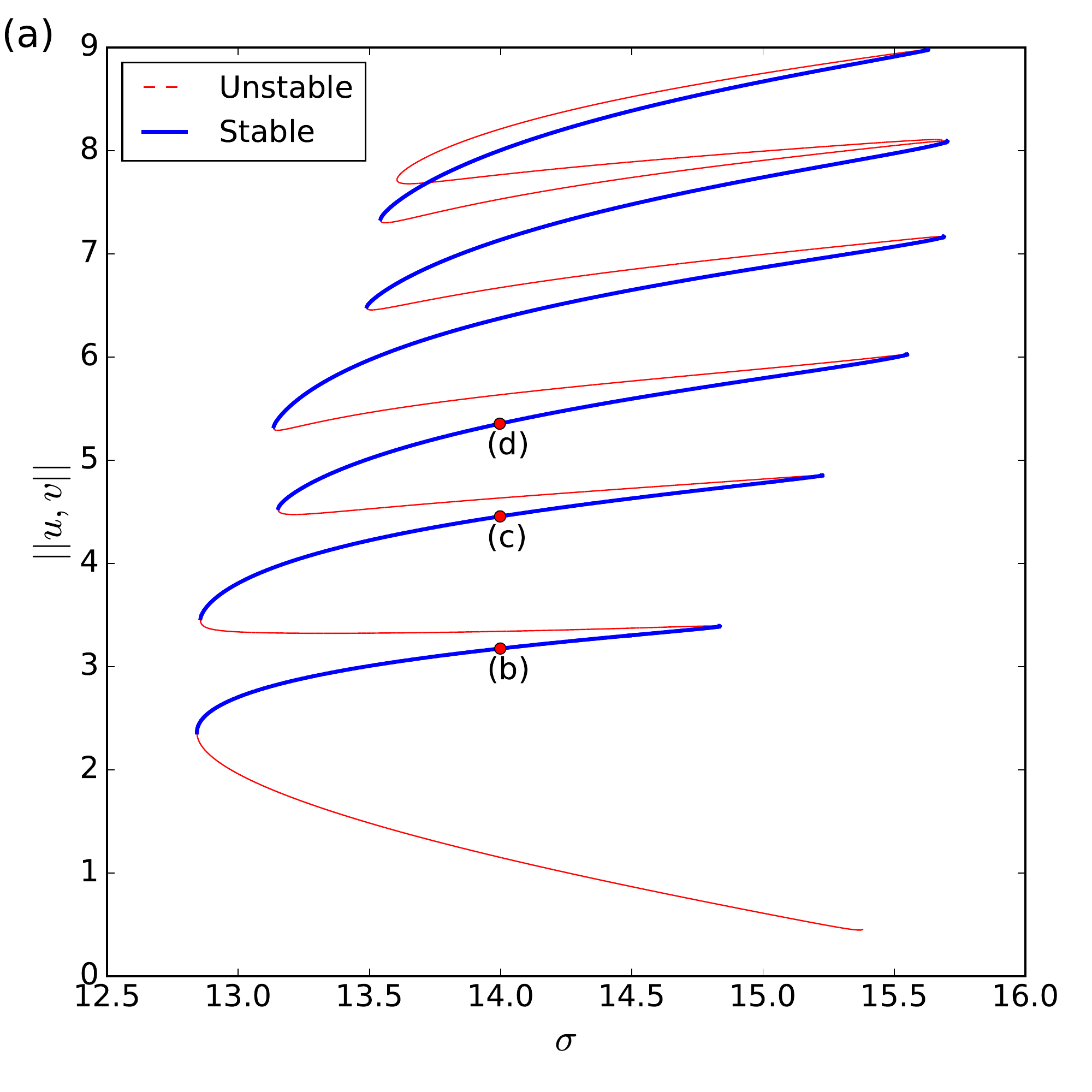}
\includegraphics[width=.49\textwidth]{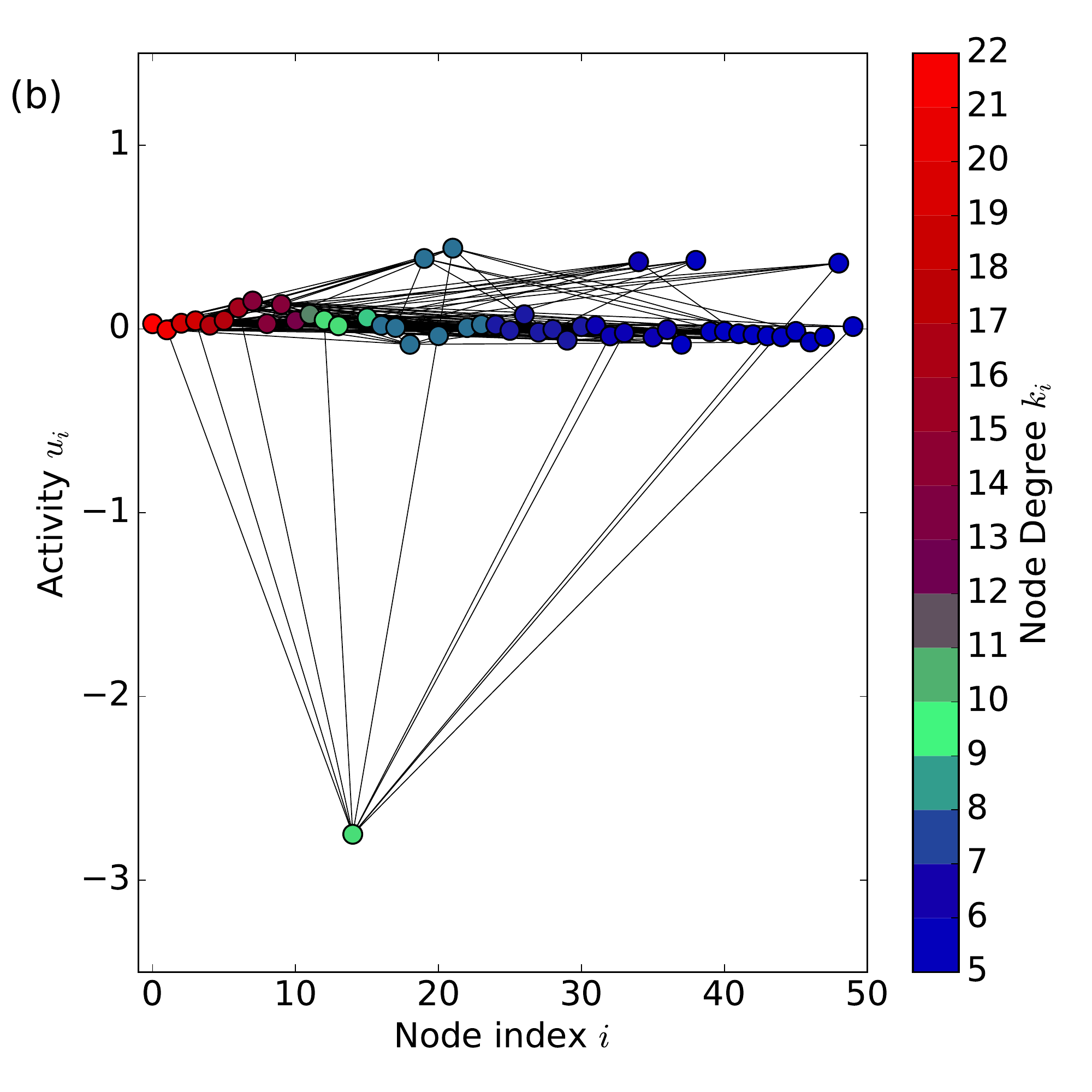}
\includegraphics[width=.49\textwidth]{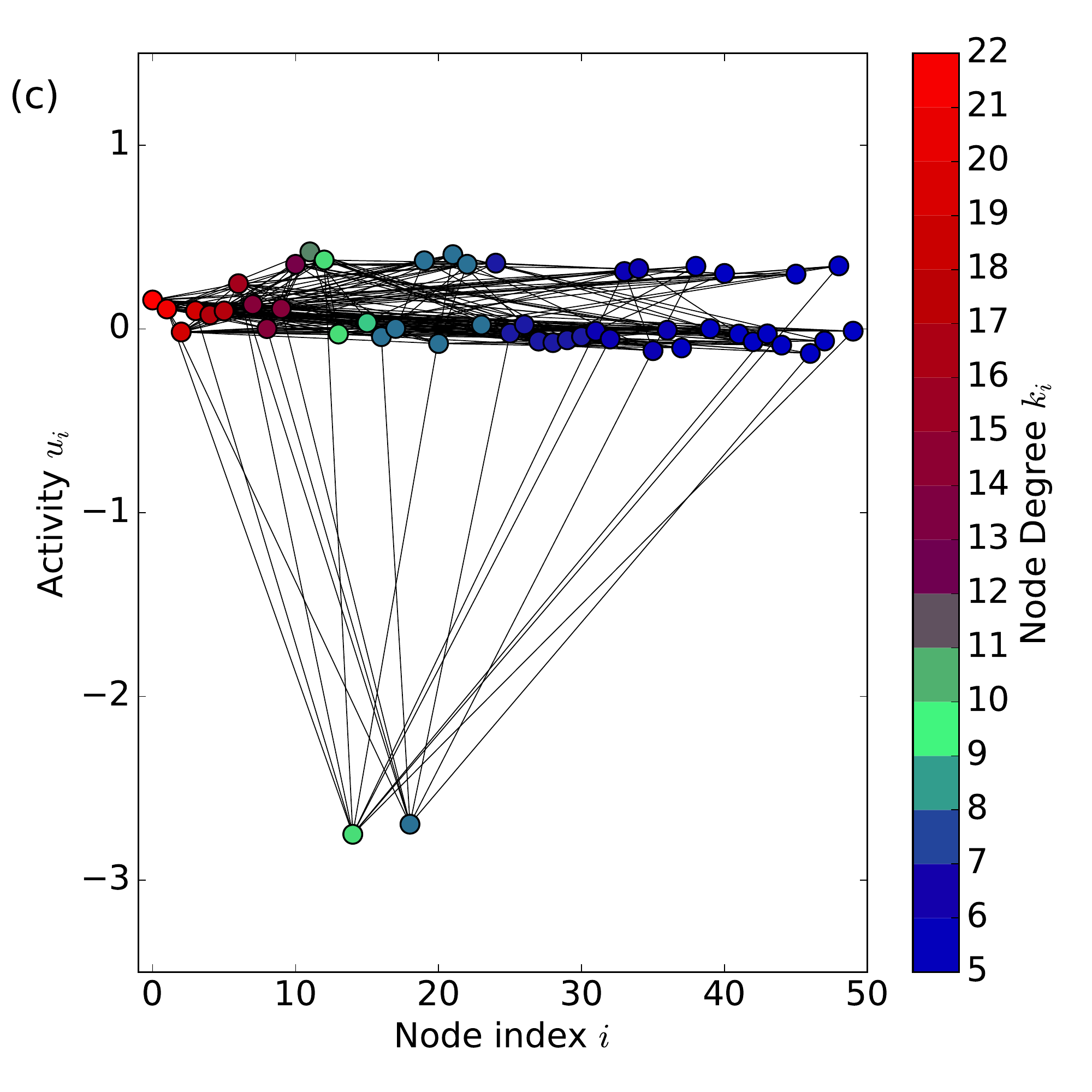}
\includegraphics[width=.49\textwidth]{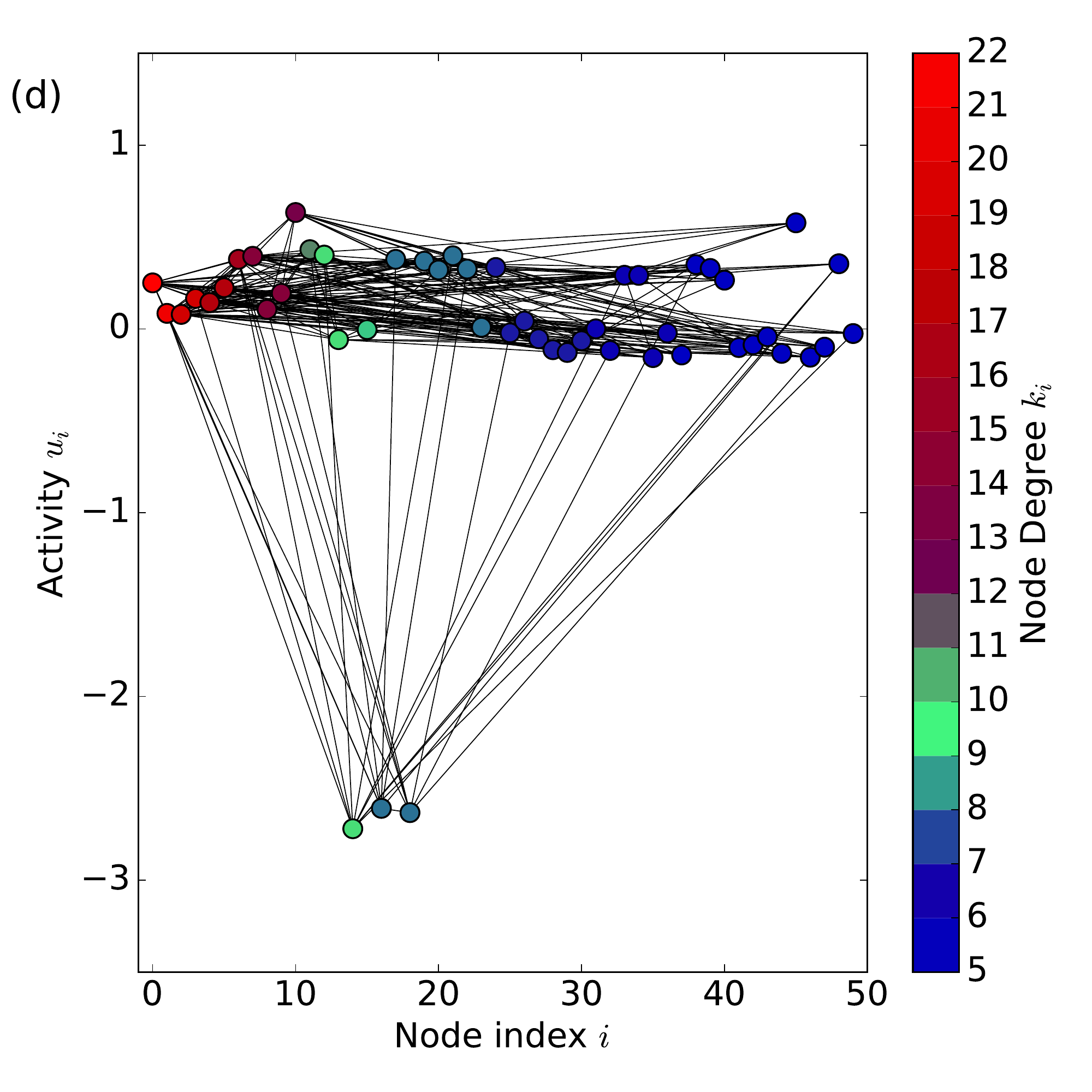}
\caption{\label{fig:snake} A snaking bifurcation (a) connecting solutions with patterns of increasing numbers of differentiated nodes. The control parameter is $\sigma$ is plotted against the magnitude of the activation vector of the two species $||\mathbf{u}, \mathbf{v}||$. Thick (blue) and thin (red) lines show stable and unstable solutions, respectively. The first three solutions are shown in  (b)--(d), with nodes ordered left--right by decreasing node degree ($d_k$ network neighbours -- also shown in colour) and the lines show the links between adjacent nodes.}
\end{figure}

As can be seen in Figure \ref{fig:snake}, for certain cases the bifurcation curve exhibits a clear {\em snaking} behaviour, with SDN solutions winding backwards and forwards in the space of the parameter $\sigma$.
Each of the bifurcation curves  fold back to the left at some point, with the associated solutions becoming unstable as another node on the network becomes differentiated from the ground-state.
This is directly analogous to the snaking bifurcations seen in regular topologies as larger patterns develop from more localised ones.  
In Figure \ref{fig:snake} each stable branch of the bifurcation curve corresponds to a different solution with different numbers of differentiated nodes, corresponding to larger values of the magnitude (L-2 Norm) of the vector displacements $||\mathbf{u}, \mathbf{v}||$, plotted on the $y$-axis.
In the example shown, with a clear snaking structure, the branches connect solutions with increasing numbers of differentiated nodes, the rest of the network remaining largely the same \footnote{This can be more clearly seen in the supplementary video \url{snaking.mov}}.
This provides further evidence that snaking is a universal phenomenon in localised pattern formation.

The snaking bifurcations can be continued further, with the bifurcation structure becoming highly complicated. 
However, for this network realisation, where the optimal degree (shown in green) is somewhere part-way between the core and periphery of the network, the bifurcations do not directly connect the patterns of ``localised'' activity to the bulk ``Turing-type'' modes.
These modes are abundant, but exist in a disconnected subset of the bifurcation turmoil.

\subsubsection{Development of large scale activation.}

Different network structures can effect different system behaviour, whereby the localised states do indeed connect to the Turing modes of full-blown system-scale activation.
For this alternative network realisation a degree of attachment of $M=10$ was used to generate the network. 
In this case the optimal degree $d_k^*=9$ nodes were instead found towards the periphery of the network, being amongst the lowest degree.

\begin{figure}[h!!]
\includegraphics[width=.49\textwidth]{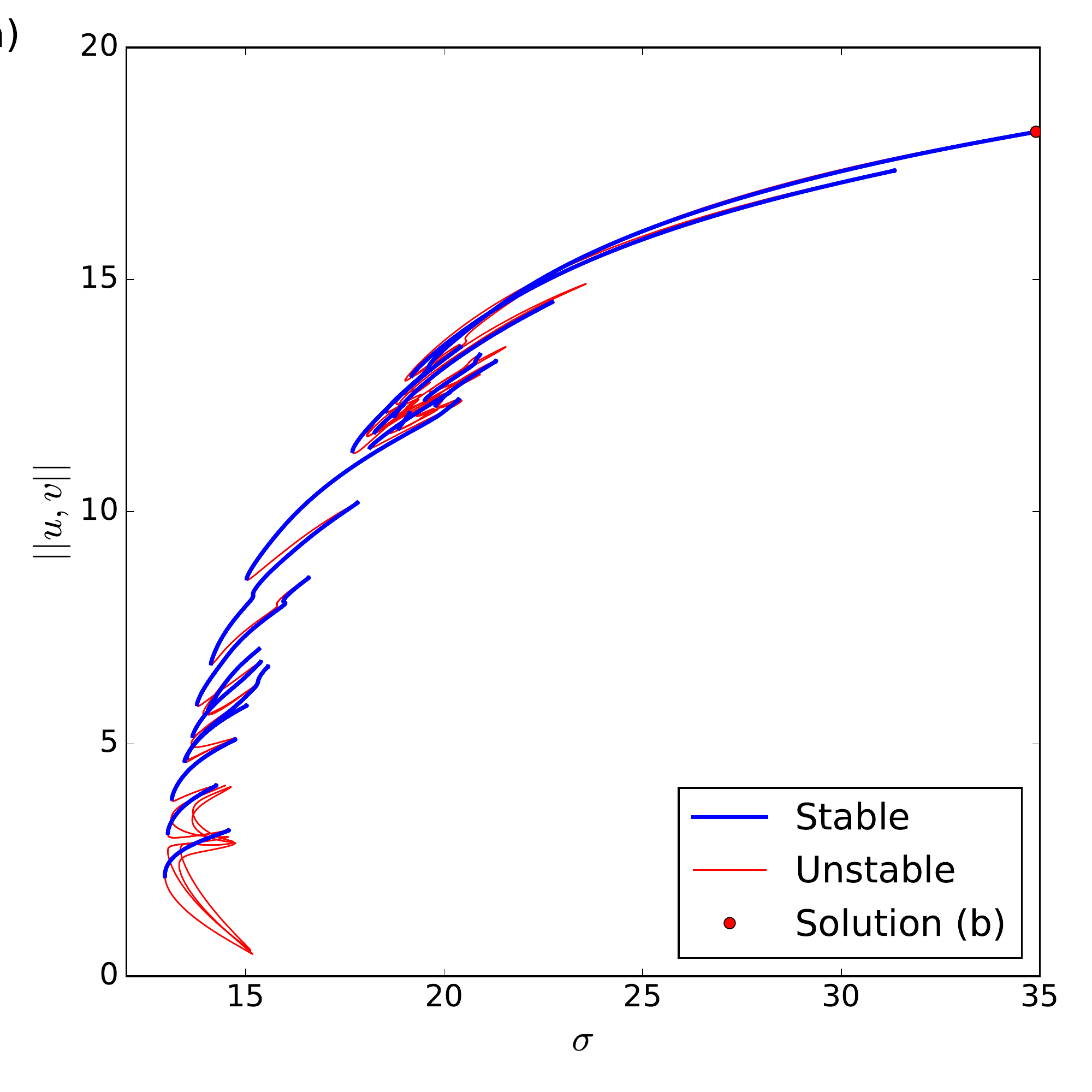}
\includegraphics[width=.49\textwidth]{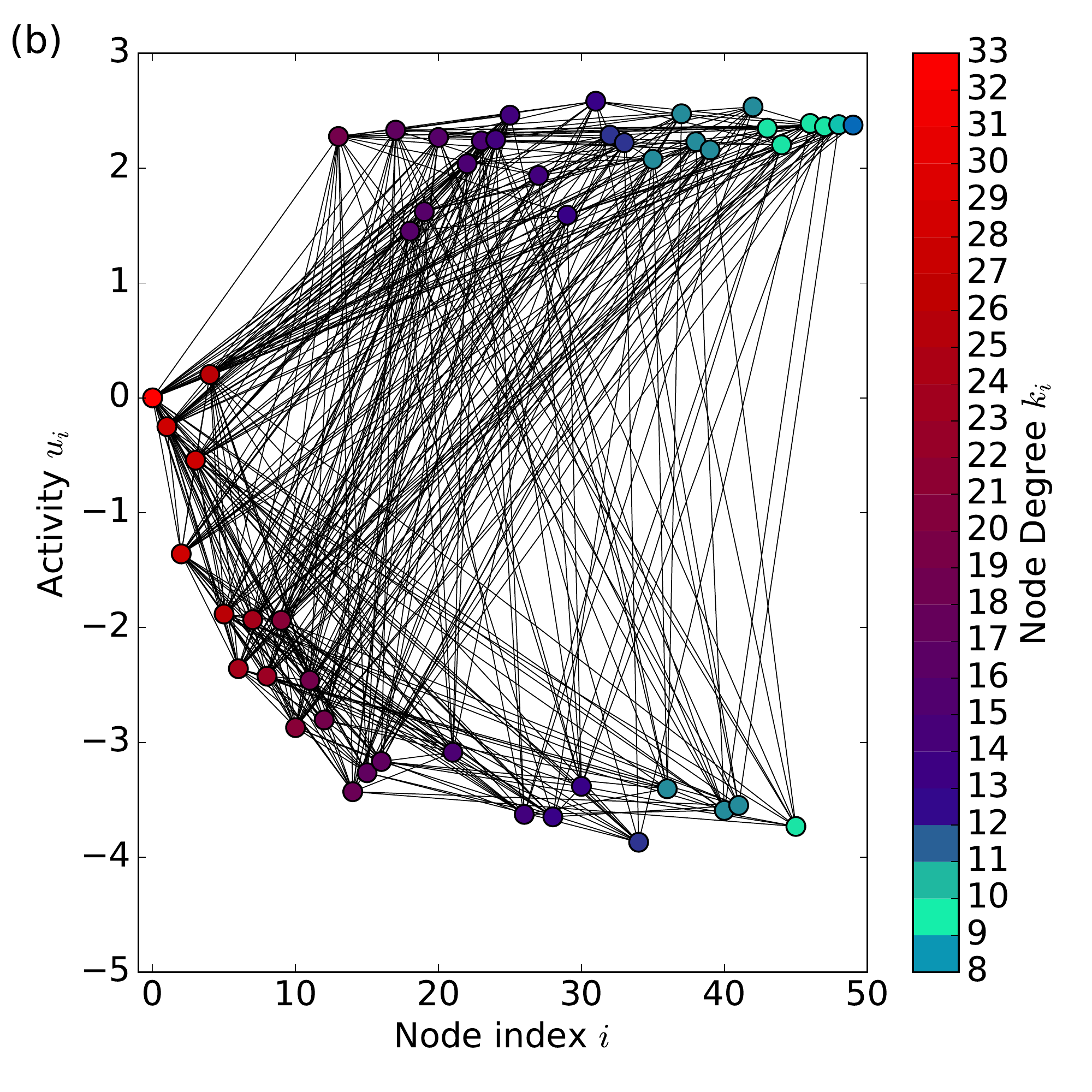}
\caption{\label{fig:toturing} Growth from localised activity to system scale {\em Turing-type} patterns in a network with {\em attachment degree} $M=10$.  (a) The snaking bifurcation diagram, with solutions at turning points numbered and (b) a bulk-mode pattern on the network nodes. The optimal degree $d_k^*=9$ nodes (shown in green) are towards the periphery of the network.}
\end{figure}

The connection between the two regimes is via a complicated series of snaking bifurcations, as shown in the example in Figure \ref{fig:toturing}, where the ``snakes'' are found to coil up and down in a cobra-like manner  as different nodes become differentiated and undifferentiated. 
This provides a clear connection between small-scale patterns and the larger bulk patterns of activity seen in the work of Nakao and Mikhailov \cite{nakao2010turing}.

The difference between this and the previous case reveals the importance of the network structure, its interaction with the reaction dynamics and the resulting behaviour of the system.
The position of the optimal degree nodes within the network strongly affects the connectedness of solutions as these are preferentially the first nodes to be activated.

\subsubsection{Disconnected Solutions}

As well as states connected to the undifferentiated {\em ground-state}, a large number of disconnected  branches were also found, similar to those seen in studies on lattice systems.
All cases where bifurcation curves included single-node differentiation of nodes with the lowest degree ($d_k=5$) were found to be of the simple {\em figure-8} isola-loop kind in this case, %(Fig.~\ref{fig:isolas} (a)).
but in other cases the bifurcations wind around in a very complicated way (see supplementary material for example plots).
Many of these complicated bifurcation curves coexist multistably, inhabiting a complicated {\em ``zoo''} of
behaviour.

It is clear that at there exist an enormous number of coexisting combinations of differentiated nodes, connected by a vast number of complicated bifurcation structures, covering a range of parameter values.
Understanding the statistical distribution of states over these values will be valuable in understanding the variety of patterns and possible configurations that can be exhibited by such systems.

\subsubsection{Multiplicity of solutions}

\begin{figure}[h]
\centering
\includegraphics[width=.49\textwidth]{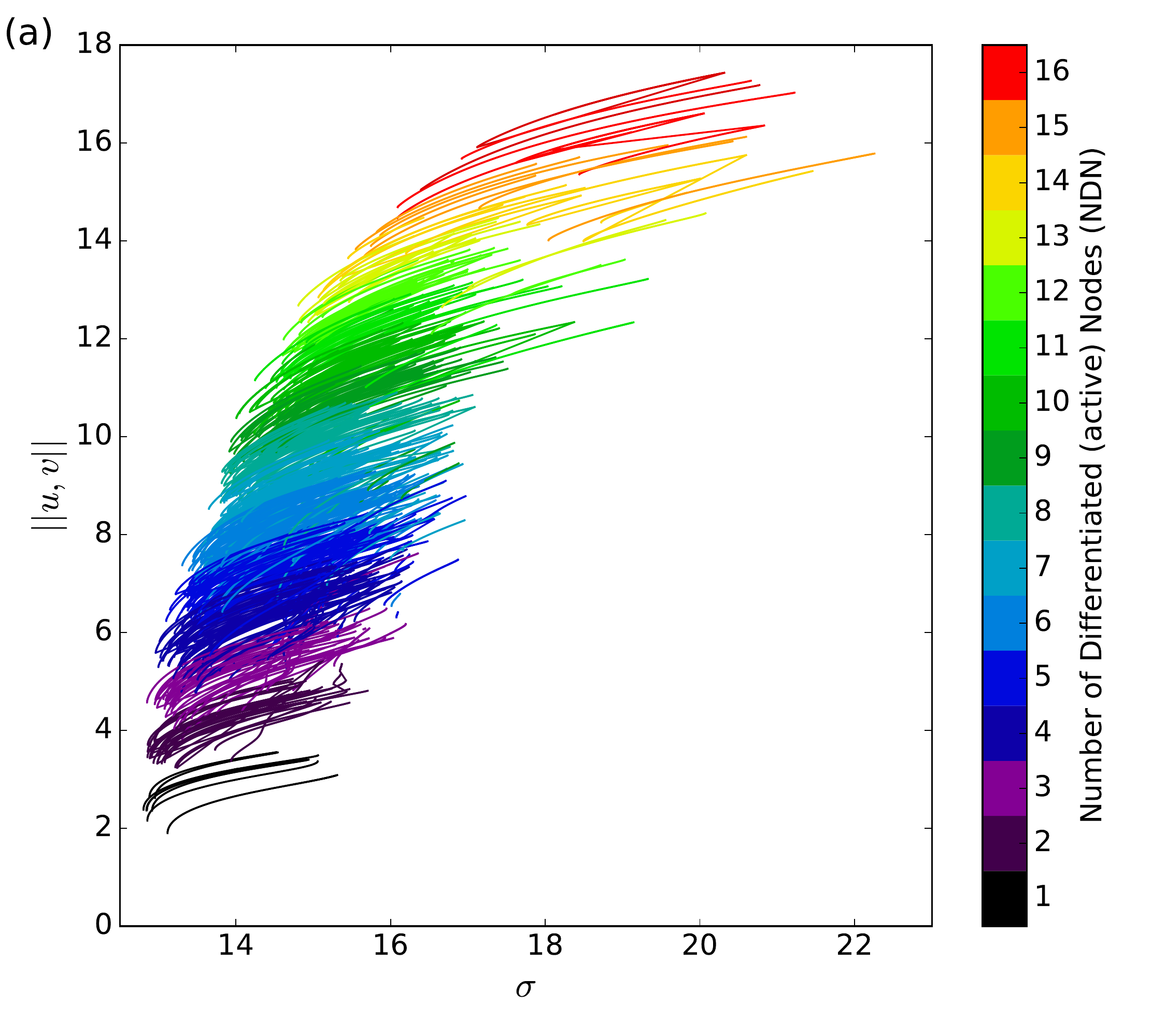}
\includegraphics[width=.49\textwidth]{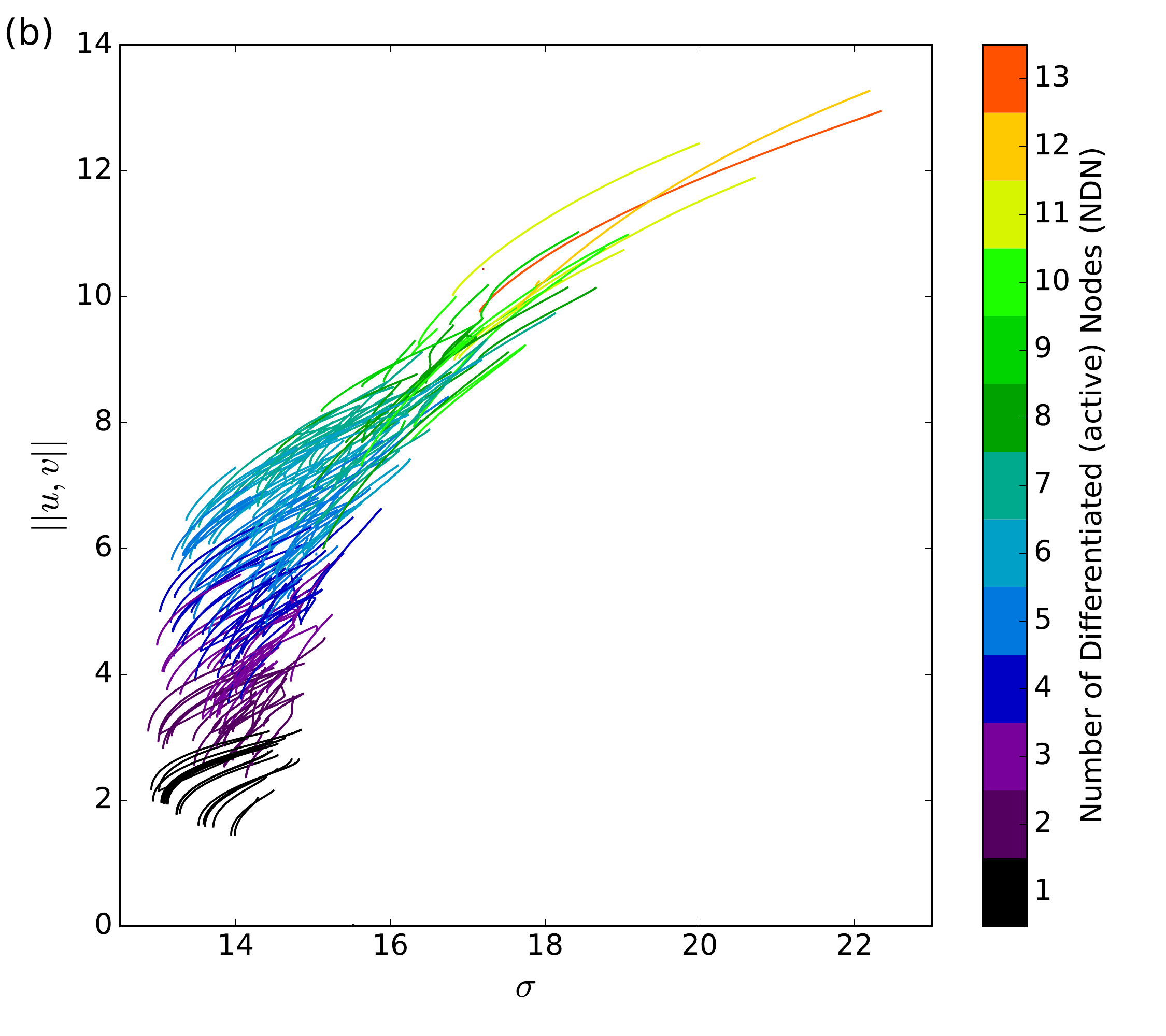}
\caption{\label{fig:mdn} Full set of solution branches at lower $\sigma$ values, showing both stable and unstable branches together. Colours indicate the number of differentiated nodes on each curve. (a) The $M=5$ networks with $d_k^*$ intermediate in position, and (b) the $M=10$ network with $d_k^*$ near the periphery.}
\end{figure}

As before, the initial SDN solutions were isolated before being continued in the parameter space, as can be seen in the lower branches of Figure \ref{fig:mdn}.

The region that these solutions occupy closely agrees with the predictions of the reduced-system approach of Wolfrum \cite{wolfrum2012turing}. 
In addition, the stable branches that appear for for the lowest value of $\sigma$ are those  with the optimum degree, as predicted from Wolfrum's work.

Each of these SDN states were then used as initial conditions for the continuation further up the snaking bifurcations into MDN solutions. 
The snaking bifurcations for all accessible states with up to 9DN are shown together in Figure \ref{fig:mdn}.

These different solutions coexist multistably at the same parameter values. 
The clear bunching seen for SDN solutions becomes more diffuse as more nodes become differentiated and the nodes interact more strongly.
Solutions are tightly bounded in the bifurcation space and contained in a region which continues to high values of $\sigma$ and bulk modes of activity. 
A notable feature of the results is that the initial ``snaking'' column, for small numbers of differentiated nodes, curves markedly to the right, in contrast to studies on regular geometries and spatial systems.
This can be explained heuristically by considering that, for each number of differentiated nodes there exists some stable branch that appears for minimal $\sigma$, which consists of solutions with differentiated nodes of optimal degree, as predicted by Wolfrum \cite{wolfrum2012turing}.
However, at some point he MDNs run out of nodes with optimal degree to be differentiated and instead another node activates, which necessarily has a higher activation $\sigma$.

Analytically determining these {\em regions of existence} for networks should be an important area for future analytical study in such complex systems. 
This should be amenable to rigorous analysis, such as extensions of the theory of {\em N-pulse solutions}  found in snaking on regular systems \cite{knobloch2008snaking, knobloch2011isolas}.

\subsubsection{Statistical density of states}

Two clear peaks can be seen in the histograms of (a) and (c), with the broader of the two  peaks appearing at higher $\sigma$ associated with bulk modes of the system.
The narrow peak at low $\sigma$ is associated with the cluster of localised solutions and the growth to extended patterns, connected via the snaking bifurcations. 
Clear differences can be seen for the two different network topologies, again highlighting the important role of network structure in organising the dynamics.
In the more highly connected $M=10$ network the bulk solutions are most abundant at lower values of the control parameter $\sigma$ and are more smoothly connected than in the case where the optimal degree is buried deep in the network.

\begin{figure}[h]
\centering
\includegraphics[width=0.49\linewidth]{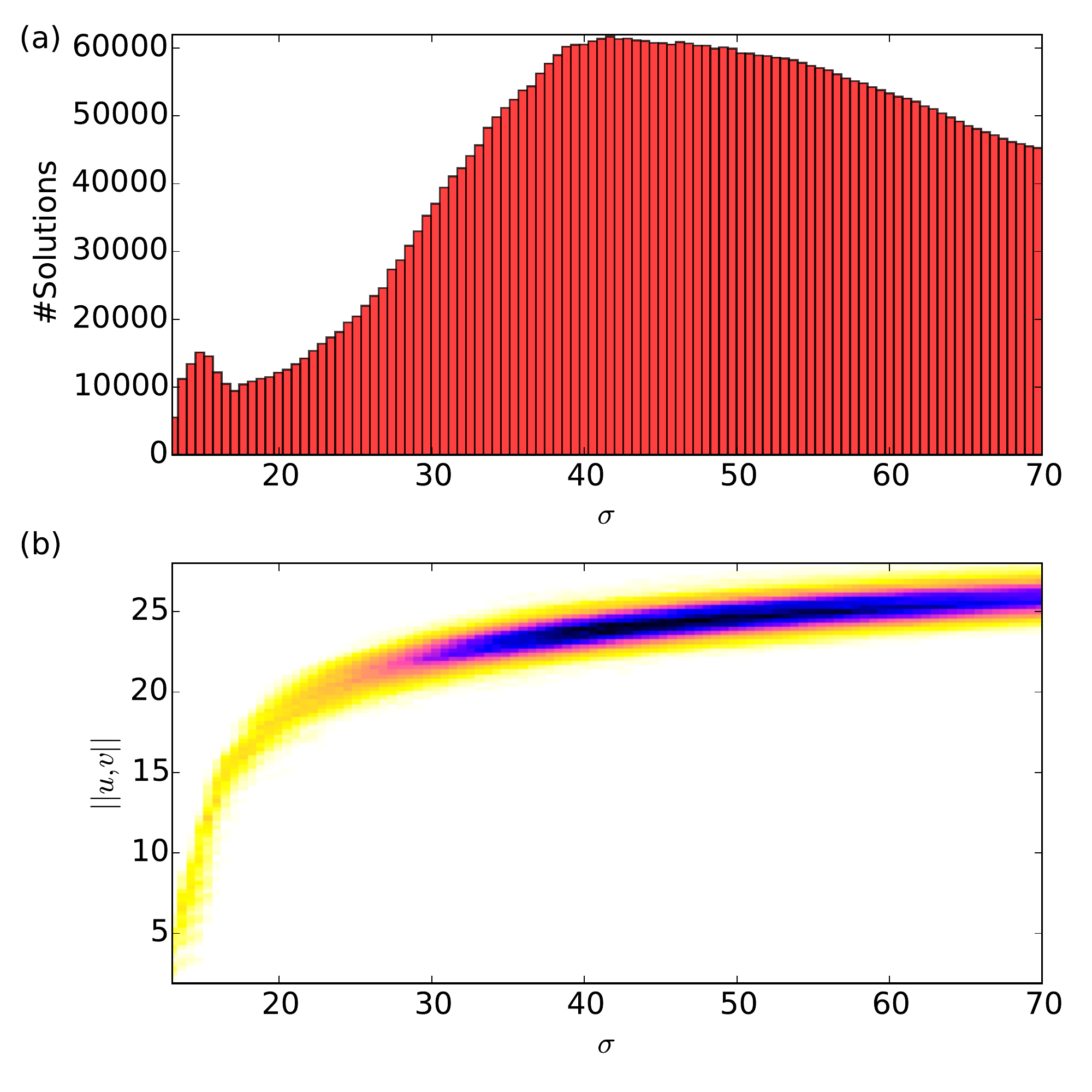}
\includegraphics[width=0.49\linewidth]{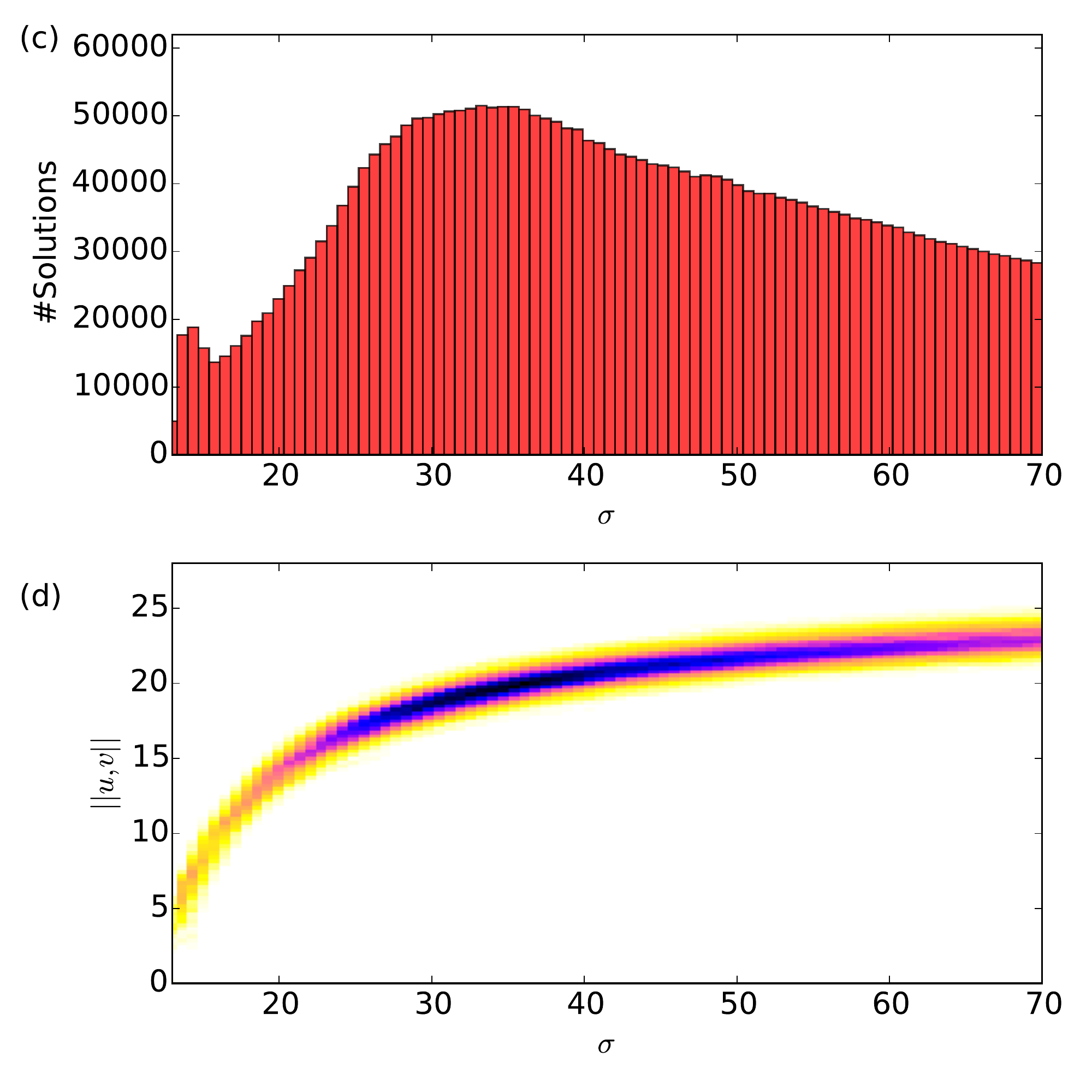}
\caption{\label{fig:hist} Statistical distributions of solutions, shown as the both histogram and 2D density plot over the bifurcation space for a range of values of the control parameter $\sigma$. (a) \& (b) are for the $M=5$ case where the small-scale patterns do not directly connect to the system-scale patterns and (c) \& (d) for the $M=10$ case where the do connect. In both cases two peaks can be seen, one for the snaking column and the other for the bulk solutions.}
\end{figure}

The region of existence described in the previous section can be most clearly seen in the two dimensional density plot shown in Figures \ref{fig:hist} (b) and (d), showing the density of bifurcation curves over a wide range of the parameter $\sigma$. 
The projection onto a one-dimensional histogram in $\sigma$ shown in (b) clearly shows the number of available system configurations at each of the parameter values.

These results reveal the potential states accessible to this system, and similar treatment could shed light on the potential configurations possible in a vast number of systems of this type arising in nature.
Examples of such systems with both activator-inhibitor kinetics as well as complex interaction topologies include: gene networks; protein species interaction networks, such as those seen in early evolution; and competition networks in other complex natural and human systems, such as social and economic systems.
 
\section{Summary and Conclusions}

In this research a reaction diffusion system on a complex network topology has been numerically investigated in detail. 
The results have revealed the transition  between the single-node solutions, studied in \cite{wolfrum2012turing}, which bifurcate from the undifferentiated state, and the fully developed patterns reported in  \cite{nakao2010turing}.
This was carried out by numerical continuation of the solutions in the parameter-space of the system and study of the solutions found along the various multistable branches found at each set of values.
The results show that snaking bifurcation, like those found in reaction--diffusion systems on regular lattice network topologies, explain the origin of and connection between the multistability of states found in previous work \cite{nakao2010turing}.

The important connection between network structure, reaction kinetics and resulting system behaviour has also been revealed, including the importance of the optimal degree nodes.

A statistical analysis of the bifurcation-structure has been used to present the multiplicity of configurations available to systems on networks. 
Understanding the density of solutions and bifurcation curves that exist within certain regions of the control parameter space is expected to reveal much about the underlying systems being modelled. 
In systems of competing chemical or biological agents the density of states could indicate the diversity of species that can exist at different values of some external environmental parameter.
Similar analogies could be made when modelling social systems, as well as many other complex systems, as competing interactions via networks.

\section*{Acknowledgements}

This work was started by, and completed in memory of, Thomas Wagenknecht.
NJM would like to thank Matthias Wolfrum and the Weierstrass Institute for Applied Analysis and Stochastics for financial support and continual advice towards this work.

\bibliography{ref}
\bibliographystyle{unsrt}

\end{document}